# Large-Scale Network Embedding in Apache Spark

Wenqing Lin
edwlin@tencent.com
Interactive Entertainment Group, Tencent
Shenzhen, Guangdong, China

## ABSTRACT

Network embedding has been widely used in social recommendation and network analysis, such as recommendation systems and anomaly detection with graphs. However, most of previous approaches cannot handle large graphs efficiently, due to that (i) computation on graphs is often costly and (ii) the size of graph or the intermediate results of vectors could be prohibitively large, rendering it difficult to be processed on a single machine. In this paper, we propose an efficient and effective distributed algorithm for network embedding on large graphs using Apache Spark, which recursively partitions a graph into several small-sized subgraphs to capture the internal and external structural information of nodes, and then computes the network embedding for each subgraph in parallel. Finally, by aggregating the outputs on all subgraphs, we obtain the embeddings of nodes in a linear cost. After that, we demonstrate in various experiments that our proposed approach is able to handle graphs with billions of edges within a few hours and is at least 4 times faster than the state-of-the-art approaches. Besides, it achieves up to 4.25% and 4.27% improvements on link prediction and node classification tasks respectively. In the end, we deploy the proposed algorithms in two online games of Tencent with the applications of friend recommendation and item recommendation, which improve the competitors by up to 91.11% in running time and up to 12.80% in the corresponding evaluation metrics.

## 1 INTRODUCTION

Graphs are so prevalent that we can model most of data as graphs naturally, e.g., social networks and knowledge graphs. Therefore, there are a large number of applications using graphs, such as recommendation systems [25] and anomaly detection [15], that extract features from graphs and utilize the machine learning models for inference tasks. However, it is difficult to manually collect useful features for each node in the graph, due to that most of graph algorithms are computationally costly. Network embedding [2, 5, 11, 13] has been a widely adopted technique for feature extraction on graphs, which computes for each node a real-valued vector of length far less than the number of nodes in the graph. Besides, it preserves the structural information of the graph, which empowers the downstream applications, such as node classification and link prediction.

However, existing network embedding algorithms often incur massive costs in the running time and memory consumption, not to mention that most of real-life graphs could be very large. For instance, node2vec [12], which is an extensively used network embedding technique, requires around one hour to process a graph with one million nodes. As such, for a sampled graph of Facebook [1] consisting of more than 700 millions of nodes, the processing time of node2vec would be at least one month, which is highly unacceptable in practice. Besides, assuming that the length of node vector is as short as 100, regardless of the huge amount of intermediate results, the memory consumption of network embedding for the sampled Facebook graph would be more than hundreds of gigabytes, which could be overwhelming to a single machine.

Recently, some techniques are proposed to compute the network embedding on large graphs. In particular, Duong et al. [9] devise a method that first divides the graph into several even-sized subgraphs by graph partitioning algorithms, then computes the network embedding for each subgraph respectively in parallel, and finally identifies the anchor nodes in the graph to integrate the embedding space from all subgraphs. In particular, the anchor nodes are chosen from the borders of the partitions, that appear in multiple partitions. However, since there could exist a large number of anchor nodes, it would be difficult to select suitable anchor nodes. Moreover, the anchor nodes are often weakly connected to the nodes in the corresponding partitions, rendering this method unable to well preserve the structural information of the graph.

In addition, Pytorch-BigGraph [17] and AliGraph [39] are the representative approaches that exploit the parameter server framework [19] to address the training of network embedding for large graph over multiple machines. Nevertheless, these approaches require a massive amount of communication between machines for random access on the graph, making them inefficient. Furthermore, there also exist some approaches tackle large graphs by graph summarization [7, 21] or graph sparsification [20, 29, 37] in a single machine, which would incur the issues where (i) they are designed sequentially making them difficult to be parallelized, or (ii) they rely on matrix manipulations that would explode the memory space when the graphs are sufficiently large.

To remedy the aforementioned issues, we devise an efficient and effective algorithm for network embedding on large graphs by exploiting the distributed share-nothing computing framework, namely Apache Spark [35], which is intensively adopted for big data processing in tremendous applications [22, 32]. In particular, we first recursively partition the graph into several small-sized subgraphs, such that (i) each subgraph can be efficiently processed on a single machine, and (ii) the subgraphs and edge cuts between them can reflect the internal and external connections of nodes to the others. Then, we perform another one round of MapReduce job [6] to compute the network embedding on each subgraph in parallel and aggregate the outputs on all subgraphs in a linear cost, resulting in the final embeddings of nodes. Compared to the previous approaches, our proposed algorithm divides the embeddings by subgraphs such that (i) the size of subgraph is relatively small which could be processed efficiently, (ii) each subgraph can be computed independently which incurs less overhead in the communication between machines, and (iii) the subgraphs carrying the nodes' internal and external connections preserve the structural information of the graph, which can be used to improve the performance of network embedding in the downstream tasks.

The contributions of this paper can be summarized as follows.

- We develop a scalable approach for network embedding, which utilizes subgraphs to well facilitate the distributed computation on billion-scale graphs.
- We devise the recursive graph partitioning algorithm, which divides a graph into even-sized subgraphs that preserve the internal and external structural information of nodes.
- We demonstrate with various experiments that the proposed algorithm outperforms the competitors by at least 4 times in running time and up to 4.25% and 4.27% in link prediction and node classification tasks respectively.
- We deploy the proposed algorithms in two games of Tencent with the applications for friend recommendation and item recommendation, and show that the propose algorithms improve the baselines by up to 91.10% in running time and up to 12.80% in the corresponding evaluation metrics.

## 2 PRELIMINARIES

In this section, we elaborate the notations and definitions that are frequently used in this paper.

### 2.1 Graphs

Let $G = (V, E)$ be a graph, where $V$ is the set of nodes and $E$ is the set of edges. We say that $u \in V$ is a *neighbor* of $v \in V$ if there exists an edge $(u, v) \in E$. Denote $N(u)$ as the set of neighbors of $u$ in $G$. Note that, for ease of explanation, we assume that the graphs in this paper are undirected, but the proposed approach can be easily extended to directed graphs, as explained in Section 4.

Given the graph $G = (V, E)$, consider a subset $V'$ of $V$, i.e., $V' \subseteq V$. The *induced subgraph* $G' = (V', E')$ on the nodes in $V'$ consists of the set $E'$ of edges whose nodes are in $V'$, i.e., $E' = \{(u, v) \in E \mid u \in V', v \in V'\}$.

A *partitioning* $\mathcal{P}$ of $G$ divides $V$ into $k$ disjoint subsets, denoted by $\mathcal{P} = \{V_1, V_2, \ldots, V_k\}$, where $k$ is a user-defined number. In other words, we have (i) $V_i \subseteq V$ for $1 \leq i \leq k$, (ii) $V_i \cap V_j = \emptyset$ for $1 \leq i < j \leq k$, and (iii) $\cup_{1 \leq i \leq k} V_i = V$. Given a node $u \in V$, let $V' \in \mathcal{P}$ be the partition where $u$ resides, denoted by $\rho(u) = V'$. We say that the neighbors in the same partition are *internal nodes*, while the others are *external nodes*.

Besides, a node $u \in V$ is a *border node* of $G$, if $u$ has at least one neighbor $v \in N(u)$ whose partition is different from the one of $u$, namely $\rho(u) \neq \rho(v)$. Let $V_b$ be the set of border nodes of $G$. The *border subgraph* $G_b$ with respect to $\mathcal{P}$ is the induced subgraph of $G$ constructed on $V_b$.

EXAMPLE 1. *Figure 1 shows a graph $G$ with 10 nodes and 12 edges. In particular, there are 3 partitions in $G$, whose nodes are colored green, yellow, and grey respectively. Then, based on the partitioning, we can construct the induced subgraphs $G_1$, $G_2$, and $G_3$ respectively, each of which consists of the nodes and edges only from a partition with the same color. For example, $G_2$ includes nodes $v_5$, $v_6$, and $v_7$, as well as the edges connecting them. On the other hand, $G$ has 4 border nodes, i.e., $v_4$, $v_5$, $v_8$, and $v_{10}$, each of which has at least one neighbor colored differently. Hence, we can construct the border subgraph $G_b$ of $G$ as the induced subgraph on the border nodes, which are connected by 3 edges.* □

### 2.2 Network Embedding

Given a graph $G = (V, E)$, network embedding is to compute for each node $u \in V$ a real-valued vector $f(u) \in \mathbb{R}^d$ of length $d$, where $d$ is a user-defined number. In addition, $f(u)$ should preserve the structural information of $G$ such that it enables the downstream applications, e.g., link prediction and node classification. Existing methods for network embedding can be roughly classified into two categories: (i) random walk based approaches, and (ii) matrix factorization based approaches.

In the random walk based approaches [12, 28], we first generate a large number of random walks for each node in the graph, and then compute the embedding of each node by maximum likelihood optimization with respect to the random walks. Denote $\Pr(v \mid f(u))$ as the likelihood of node $v$ with respect to node $u$ given the embedding $f(u)$ of $u$. Besides, following [12], denote $N_S(u)$ as the set of *sampled neighbors* of $u$ obtained by the *sampling strategy*, e.g., random walks starting from $u$. The objective of these approaches can be summarized as to maximize the log likelihood of neighbors in the graph, i.e.,

$$O_r = \log \prod_{u \in V} \prod_{v \in N_S(u)} \Pr(v \mid f(u)). \quad (1)$$

On the other hand, the matrix factorization based approaches [26, 30] model the network embedding problem as the matrix factorization problem by approximating the dot product of two nodes' vectors with their similarity in the graph. Hence, it is conceptually to minimize

$$O_m = \sum_{u \in V} \sum_{v \in V} (f(u)^\top \cdot f(v) - A_{u,v})^2, \quad (2)$$

where $A$ is the similarity matrix of nodes in the graph $G$, e.g., the adjacency matrix of $G$, and $A_{u,v}$ denotes the similarity of nodes $u$ and $v$ in $G$.

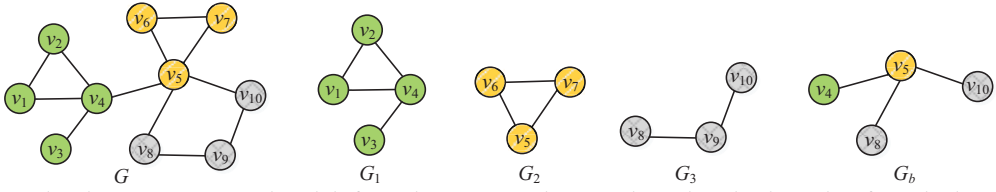

Figure 1: A graph $G$ has 3 partitions colored differently. $G_1$, $G_2$, and $G_3$ are the induced subgraphs of $G$, which are induced on the partitions respectively. $G_b$ is the border subgraph of $G$ induced on the boarder nodes.

## 2.3 Distributed Computing using Apache Spark

However, it would be impractical to compute the network embedding on a single machine. The reasons are two-fold: (i) The graph could be too large to fit in the memory of a single machine, not to mention that each node is associated with a few real-valued vectors of sufficiently large length; (ii) The computational cost of network embedding could be extremely high, due to a massive amount of random graph traversal or matrix manipulations.

To address the above issues, we develop a distributed algorithm by exploiting subgraphs to compute the network embedding on large graphs. In particular, we implement our algorithms in Apache Spark, which is a share-nothing distributed computing platform based on the MapReduce paradigm [6, 24, 33] and widely used for big data processing in many applications [22, 32]. In a MapReduce job, there are mainly three consequent phases: (i) The Map phase applies a map function on each data tuple consisting of a key-value pair, which takes as input a key-value pair and outputs new key-value pairs. (ii) Then, it is the Shuffle phase where all new key-value pairs are shuffled among machines, such that the pairs with the same key are sent to the same machine and aggregated together. (iii) Finally, the Reduce phase exploits a reduce function on the key-value pairs sharing the same key, which outputs a new key-value pair, leading to the result of this MapReduce job. Note that, a distributed algorithm on Apache Spark could consist of several consecutive MapReduce jobs, each of which could be fed with the results from the prior jobs.

## 3 NETWORK EMBEDDING VIA SUBGRAPHS

Before introducing the proposed distributed algorithms for network embedding, we present the intuitions behind it. As nodes in a graph are more likely to be connected with the ones in a close distance [1, 27], the local structures should be critical to the network embedding of the graph. In light of that, we show that the network embedding of a graph is dividable, which can be approximated with the ones of its subgraphs.

To explain, given a graph $G = (V, E)$ and a number $k$, we divide $G$ into $k$ partitions, resulting in $\mathcal{P} = \{V_1, V_2, \ldots, V_k\}$. For $1 \le i \le k$, we denote the subgraph induced on $V_i$ as $G_i$.

Consider Eq. 1 in the random walk based approaches. For each node $u$ in $G$, we can separate its neighbors $v$ by inspecting whether $u$ and $v$ are in the same partition, i.e., $\rho(u) = \rho(v)$.

Hence, we can modify Eq. 1 as

$$O_r = \log \prod_{u \in V} \prod_{\substack{v \in N_S(u) \\ \rho(u)=\rho(v)}} \Pr(v \mid f(u)) + \log \prod_{u \in V} \prod_{\substack{v \in N_S(u) \\ \rho(u) \ne \rho(v)}} \Pr(v \mid f(u)).$$

Let $O_p$ and $O_{np}$ be the first and second parts of the right hand side formula respectively. Note that, $O_p$ focuses on the nodes in the same partition. In other words, it considers only the probability of nodes $u$ and $v$, which are both in the subgraph $G' \subseteq G$ induced on a partition $V' \in \mathcal{P}$. As a result, we can approximate $O_p$ with the probabilities of nodes in each subgraph $G_i$ of $G$ individually, where $1 \le i \le k$. Given a network embedding $f_i$, denote the log likelihood of nodes in $G_i = (V_i, E_i)$ as

$$O(f_i) = \log \prod_{u \in V_i} \prod_{v \in N_{S_i}(u)} \Pr(v \mid f_i(u)),$$

where $N_{S_i}(u)$ is the set of sampled neighbors of $u$ in $V_i$. Hence, we can approximate $O_p$ using $\sum_{i=1}^{k} O(f_i)$. Note that, since $\mathcal{P}$ is a node disjoint partitioning, each node $u$ in $G$ would be covered by only a $f_i$ where $1 \le i \le k$.

Now consider $O_{np}$, which concentrates on the probabilities of nodes $u$ and $v$ that are not in the same partition, i.e., $\rho(u) \ne \rho(v)$. Denote the log likelihood of sampled neighbors of nodes in $G_b$ as

$$O(f_b) = \log \prod_{u \in V_b} \prod_{v \in N_{S_b}(u)} \Pr(v \mid f_b(u)),$$

where $N_{S_b}(u)$ is the set of sampled neighbors of $u$ in $V_b$. Hence, we have $O(f_b)$ as the approximation of $O_{np}$.

Putting all together, to maximize Eq. 1, we can perform the maximum likelihood optimization for each induced subgraph and border subgraph of $G$ respectively, which approximates $O_r$.

On the other hand, for the matrix factorization based approaches, Eq. 2 can be expanded in a similar way:

$$O_m = \sum_{i=1}^{k} \sum_{u \in V_i} \sum_{v \in V_i} (f(u)^\top \cdot f(v) - A_{u,v})^2 + \sum_{u \in V} \sum_{\substack{v \in V \\ \rho(v) \ne \rho(u)}} (f(u)^\top \cdot f(v) - A_{u,v})^2.$$

Consequently, $O_m$ can also be approximated by independently computing the network embedding for the subgraphs of $G$, which can be analyzed similar to the one of random walk based approaches.

It is worthy noting that although the proposed approach approximates network embedding via subgraphs, its performance can still outperform the other implementations, as shown in Section 5. This is because the proposed approach differentiates the internal and external information between nodes with the induced and border subgraphs respectively, which together augment the structural information preserved by the produced network embeddings of nodes.

## 4 DISTRIBUTED NETWORK EMBEDDING

To compute the network embedding for large graphs on multiple machines, one straightforward approach is to exploit the parameter server [19] which is widely used for machine learning on big data. In the architecture of parameter server, the workers are falling into two types, namely clients and servers, where the clients can easily access the data stored in the servers with push/pull operations. However, due to that the training of network embedding would require random access to different portions of the graph, it would incur an extremely high cost in the communication between clients and servers, rendering this approach highly inefficient.

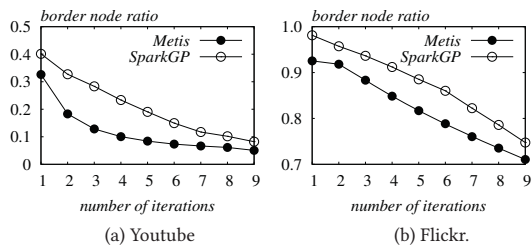

(a) Youtube     (b) Flickr.

**Figure 2: Border node ratio, defined as the number of border nodes over the number of nodes in the given graph.**

To address this issue, we propose an approach based on Apache Spark running in three phases, as follows.

- First, we recursively divide the graph $G$ into several even-sized subgraphs $G'$. In particular, a graph $G$ is partitioned to generate several subgraphs, as well as a border graph which will be further partitioned if its size is still large. As such, the embedding of a node in $G$ is constructed as the combination of the embeddings in the subgraphs and the border subgraph respectively.
- Then, for each subgraph $G'$ produced by the recursive graph partitioning algorithm, we compute its network embedding independently. As such, we are able to compute the network embedding for all subgraphs efficiently with the distributed computing framework, and avoid the costly communication among different machines.
- Finally, the embeddings of each node in all subgraphs are fused, resulting in the embedding of each node in $G$, as explained in Section 3. By this means, we are able to take into account both the internal and external connections of nodes for the network embedding of $G$.

In the sequel, we elaborate the details of each phase in the proposed approach.

### 4.1 Recursive Graph Partitioning

There exist a number of algorithms for graph partitioning, such as Metis [16]. However, most of them are tailored for a single machine, rendering them unsuitable for processing large graphs with distributed computing. To facilitate the graph partitioning for large graphs in Spark, we adopt the approach [3], denoted by *SparkGP*, which exploits the greedy strategy by swapping nodes iteratively to minimize the cuts among different partitions.

Recall that, given a graph $G$, we divide $G$ into $k$ partitions, which lead to $k$ induced subgraphs and a border subgraph $G_b$. After that, we compute the network embedding for all subgraphs in parallel, each of which is processed in a single machine. However, the size of the border subgraph could be still large that are too difficult to be processed by a single machine (see Figure 2).

To remedy this issue, we propose to partition the graph $G$ recursively. Specifically, we first partition $G$ into $k_1$ induced subgraphs,

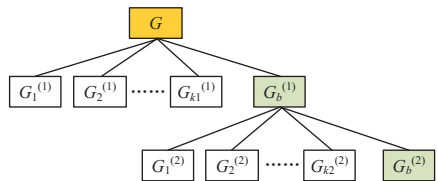

**Figure 3: Recursive graph partitioning on $G$ with 2 iterations.**

denoted by $G_1^{(1)}, G_2^{(1)}, \ldots, G_{k_1}^{(1)}$ and a border subgraph $G_b^{(1)}$, where $k_1$ equals the $k$, i.e., the initial number of graph partitions. If the size of $G_b^{(1)}$ is still large, we perform the second round of graph partitioning on $G_b^{(1)}$, resulting in another $k_2$ induced subgraphs $G_1^{(2)}, G_2^{(2)}, \ldots, G_{k_2}^{(2)}$ and the border subgraph $G_b^{(2)}$. Note that, $k_2$ is not necessarily equal to $k_1$, due to that the size of $G_b^{(1)}$ is often smaller than the size of $G$. In particular, to make sure that the induced subgraphs of $G_b^{(1)}$ are of similar size as the ones of $G$, we have $k_2 = \frac{|V_b^{(1)}|}{|V|} k_1$, where $V_b^{(1)}$ is the set of nodes of $G_b^{(1)}$. After that, we recursively partition the border subgraph $G_b^{(2)}$, until (i) the border subgraph $G_b^{(j)}$ in the $j$-th iteration can fit in the memory of a single machine, or (ii) the number of iterations $\gamma$ has reached a user-defined maximum number, which will be discussed later. As a result, the recursive graph partitioning algorithm divides the graph $G$ into several induced subgraphs generated in a few iterations, as well as a border subgraph in the last iteration.

Figure 3 shows a tree to illustrate the recursive graph partitioning on the graph $G$ with 2 iterations. If a graph is partitioned due to its sufficiently large size, its subgraphs are denoted as its children on the tree. Therefore, all leaves on the tree are the resulting subgraphs computed by the recursive graph partitioning algorithm.

To demonstrate the necessity of recursive graph partitioning, we run Metis and SparkGP on the *Youtube* and *Flickr* graphs respectively (see Table 1). Figure 2 shows the number of border nodes over the number of nodes in two graphs respectively. Initially, more than 30% of nodes in *Youtube* graph are border nodes, and *Flickr* graph has more than 90% of nodes that are border nodes. As we recursively partition the border subgraphs, the number of border nodes decreases greatly for both Metis and SparkGP.

Note that, there are two parameters $k$ and $\gamma$ in the proposed approach, whose setting would affect the performance of the proposed algorithm. In the following, we show that $k$ and $\gamma$ can be configured automatically.

Consider $k$. If $k$ is too large, while the computation on subgraphs would be fast due to their small size, the number of induced subgraphs becomes a lot, leading to a border subgraph of larger size. Hence, the running time of recursive graph partitioning would increase, and the final embedding for a large $k$ would be lack of effectiveness, as shown in Section 5. On the other hand, if $k$ is too small, the subgraphs would be too large to be processed efficiently on a single machine. Therefore, the choice of $k$ should take into account the size of available memory $M$ on a single machine. Given a graph $G = (V, E)$, let $n$ be the number of nodes in $V$ and $\Delta$ is the average size of memory for storing the set of neighbors for a node in $G$. Since the graph partition algorithm divides $G$ into several even-sized subgraphs, the number of nodes in a subgraph is around $\frac{n}{k}$. Hence, we have $\frac{n}{k}\Delta \leq M$ such that the size of a subgraph is less than the memory size of a single machine. In other words, we can estimate $k = \lceil \frac{n\Delta}{M} \rceil$.

Table 1: Datasets.

| Dataset | #Nodes | #Edges | #Labels |
|---|---|---|---|
| Blog[1] | 10,312 | 333,983 | 39 |
| Flickr[1] | 80,513 | 5,899,882 | 195 |
| Youtube[1] (YT) | 1,138,499 | 2,990,443 | 47 |
| Spammers[2] (SP) | 5,321,961 | 546,799,071 | 2 |
| UK2002[3] (UK) | 18,484,053 | 298,113,385 | 0 |
| Twitter[4] | 41,652,230 | 1,468,365,182 | 0 |
| Friendster[5] (FS) | 68,349,466 | 2,586,147,869 | 0 |

On the other hand, the selection of the number of iterations $\gamma$ should consider the length of embeddings of the nodes, since (i) the embedding computed by an iteration consists of the embeddings on induced subgraphs and the border subgraph, and (ii) the length of embeddings in the later iteration is smaller than that of the prior ones. To make the embedding of the border subgraph on the $\gamma$-th iteration useful, its vector length should be more than a threshold, e.g., 10. Assume that the fraction of border subgraph's embedding in the given iteration is $\delta$, where $0 < \delta < 1$. One way to decide $\delta$ is by considering both the sizes of $G$ and $G_b$, i.e., $\delta = \frac{|V_b|}{|V|+|V_b|}$. Hence, we have $d\delta^\gamma \geq 10$, where $d$ is the length of final embeddings in $G$, which results in an upper bound of $\gamma$, i.e., $\gamma \leq \log_{1/\delta}^{d/10}$.

### 4.2 Processing on Subgraphs

Given the subgraphs generated by the recursive graph partitioning algorithm running in $\gamma$ iterations, we compute the embedding of all nodes in $G$ by one MapReduce job. Specifically, in the Map phase, the network embedding of each subgraph $G^{(j)}$ is computed in a machine independently by any existing network embedding techniques, such as node2vec [12], where $1 \leq j \leq \gamma$. Then, for each subgraph $G^{(j)}$, we emit all nodes and their embeddings in $G^{(j)}$ as key-value pairs, where the key is the node $v$ in $G^{(j)}$ and the value consists of three parts: (i) the iteration number $j$, (ii) the identifier $q$ to inspect whether $G^{(j)}$ is the border subgraph, and (iii) the embedding of $v$ in $G^{(j)}$. Let $q = 1$ be the identifier for the border subgraph, and $q = 0$ for the induced subgraph. After that, all key-value pairs are shuffled among machines and aggregated, such that the embeddings of the same node $v$ are sent to the same machine. Finally, all embeddings of node $v$ are fused according to the iteration number $j$ and the identifier $q$, leading to the final embedding of $v$, which will be explained in Section 4.3.

Note that, to facilitate the computation of network embedding on the subgraphs, we require the embedding length as one of the hyperparameters. To achieve that, we calculate the embedding length for each subgraph recursively. For example, the border subgraph $G_b^{(1)}$ of $G$ has an embedding length of $\lceil \delta d \rceil$, where $d$ is the embedding length of nodes in $G$ and $\delta$ is the fraction of border subgraph's embedding in the final embedding, as discussed in Section 4.1. Hence, each induced subgraph of $G$ has the embedding length equal to $d - \lceil \delta d \rceil$. If $G_b^{(1)}$ is still large for a single machine, we recursively partition its embedding length $\lceil \delta d \rceil$ for the subsequent subgraphs with the fraction $\delta$.

[1] http://socialcomputing.asu.edu/pages/datasets
[2] https://linqs-data.soe.ucsc.edu/public/social_spammer
[3] http://law.di.unimi.it/webdata/uk-2002
[4] http://konect.uni-koblenz.de/networks/twitter
[5] http://konect.uni-koblenz.de/networks/friendster

As aforementioned, the recursive graph partitioning algorithm results in only one border subgraph, i.e., $q = 1$, which is generated in the last iteration. Besides, since the induced subgraphs produced in each iteration are node disjointed, the embedding length of the induced subgraphs in the same iteration are the same. Therefore, we can exploit the iteration number $j$ and its identifier $q$ to identify the embedding length for the subgraph $G^{(j)}$, denoted as $\ell(j, q)$. Hence, we have

$$\ell(j, q) = \begin{cases} 0, & \text{if } q = 1 \text{ and } j < \gamma; \\ \lceil \delta^j d \rceil, & \text{if } q = 1 \text{ and } j = \gamma; \\ \lceil \delta^{j-1} d \rceil - \lceil \delta^j d \rceil, & \text{otherwise.} \end{cases}$$

In other words, the sum of all $\ell(j, q)$ equals $d$, i.e.,

$$\sum_{1 \leq j \leq \gamma \text{ and } q \in \{0,1\}} \ell(j, q) = d.$$

### 4.3 Embedding Fusion

Given the key-value pairs generated from all subgraphs, the ones of the same node $v$ are aggregated as a set, denoted by $B(v)$. Based on that, we inspect all embeddings in $B(v)$ once to obtain the final embedding $f(v)$ of $v$ in $G$.

To explain that, we first calculate the starting position $s(j, q)$ in $f(v)$ for each embedding $f_{j,q}(v) \in B(v)$ with respect to the subgraph $G^{(j)}$. Specifically, $s(j, q)$ is equal to the sum of all $\ell(j', q')$ where $j' \leq (j+q)$ and $q' = 0$, i.e., we have $s(j, q) = \sum_{j' \leq (j+q)} \ell(j', 0)$. Then, we create for the node $v$ a length-$d$ vector $f(v)$ initialized with zero values. After that, for each embedding $f_{j,q}(v) \in B(v)$, we replace with $f_{j,q}(v)$ the values in $f(v)$ whose positions start from $s(j, q)$. Finally, we obtain the final embedding $f(v)$ of $v$ after all replacements are completed. As there is only once scan of the embedding in $B(v)$, the complexity of embedding fusion is linear to the length of embedding, i.e., $O(d)$.

## 5 EXPERIMENTAL EVALUATIONS

To evaluate the performance of the proposed approach, named *DistNE*, we adopt 7 datasets that are widely used in the literature, as shown in Table 1. In particular, there are 4 datasets each of which has multiple labels on their nodes, and 2 datasets containing billions of edges, which can not be handled by most of the previous algorithms, as shown in Section 5.1.

Based on the datasets, we evaluate the performance of DistNE against 7 state-of-the-art methods (see Section 7), i.e., node2vec [12], NetSMF [29], ProNE [37], SepNE [20], MILE [21], ParNE [9], and Pytorch-BigGraph (PBG) [17]. Note that, while we exploit node2vec in the implementation of DistNE, the other single-machine network embedding algorithms can also be used in DistNE.

By default, we set the length of embedding $d = 128$. Besides, for each previous algorithm, we adopt the parameters provided by their authors as the default ones. We run these algorithms on an in-house cluster with 51 machines installed with CentOS 6.4, each of which has an Intel(R) Xeon(R) CPU E5-2670 CPU with 2.30GHz and 16GB memory. For those running on single machine, we randomly choose one machine from the cluster. In each experiment, we perform each algorithm 5 times and report the average reading.

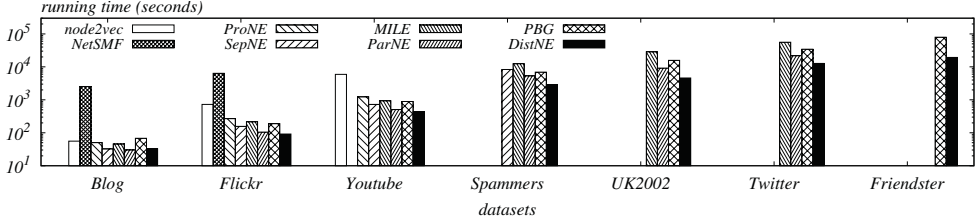

Figure 4: Running time on all datasets.

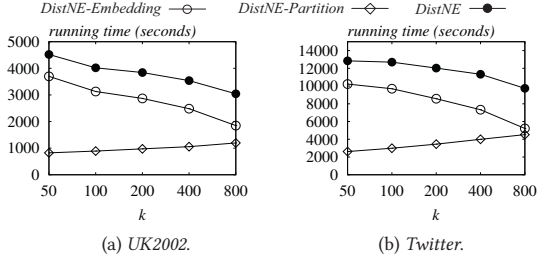

(a) *UK2002*.  (b) *Twitter*.

Figure 5: Varying $k$.

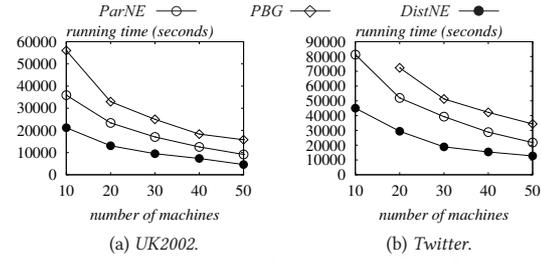

(a) *UK2002*.  (b) *Twitter*.

Figure 6: Varying the number of machines.

### 5.1 Experiments on Efficiency

First, we evaluate the running time of each algorithm on all datasets, where 5 of them have more than one million of nodes respectively. We omit the result of an algorithm on a dataset, if (i) the algorithm cannot handle the dataset with the issue of memory overflow, or (ii) the algorithm cannot finish within 24 hours. As shown in Figure 4, DistNE is able to handle all datasets, and outperforms the all competitors on all datasets. In particular, DistNE is more than an order of magnitude faster than node2vec on the *Youtube* dataset, and more than 7 times faster than MILE on the *UK2002* dataset, which demonstrates the superiority of the distributed computing approach compared with the single-machine ones. Besides, the results of the matrix based approaches, i.e., NetSMF, ProNE, and SepNE, are missing on most of the large graphs, such as Twitter and Friendster. This is because of the high cost in matrix manipulations that cause the issue of memory overflow, rendering them difficult to handle large graphs. Furthermore, DistNE achieves more than 4 times faster than PBG on the *Friendster* dataset, since PBG incurs lots of overheads in the communication between clients and servers. DistNE is more efficient compared with PBG, since DistNE employs the share-nothing distributed computing framework and allows the embedding to be computed via subgraphs of sufficiently small size in a divide-and-conquer manner. Note that, the result of ParNE on the largest dataset, namely *Friendster*, is missing, since it cannot finish within 24 hours, which is caused by the intensive cost in embedding replacements, as explained in Section 7.

### 5.2 Experiments on Parameter Sensitivities

We now study the parameters, i.e., the number of machines and the number $k$ of partitions, that would affect the performance of the proposed distributed computing algorithm.

Figure 5 depicts the running time of DistNE on the datasets *UK2002* and *Twitter* respectively by varying $k$ from 50 to 800. In particular, we decompose the running time of DistNE into two parts: (i) the one for recursive graph partitioning, denoted by *DistNE-Partition*, and (ii) the one for computing network embedding on subgraphs, denoted by *DistNE-Embedding*. As shown, the running time of DistNE decreases when $k$ increases, since the size of subgraphs becomes smaller, leading to the large decrease in the running time of network embedding on the subgraphs. Note that, while the running time of recursive graph partitioning increases slightly when $k$ increases, it is not the dominating one.

Figure 6 shows the running time of the distributed computing solutions, i.e., ParNE, PBG, and DistNE, on the datasets *UK2002* and *Twitter* respectively by varying the number of machines. Specifically, we set the number of machines as the value in the range from 10 to 50. As we can see, when we increase the number of machines, the running time of all algorithms decreases, due to the power of parallelism. Besides, our proposed approach DistNE consistently outperforms the others in all cases. Note that, the speedup is not linearly proportional to the number of machine, due to the overhead of communication between machines.

### 5.3 Experiments on Link Prediction

In this set of experiments, we evaluate the performance of all algorithms on all datasets by running the task of link prediction. In order to perform the link prediction task, for each dataset $G = (V, E)$, we randomly remove $\lfloor \alpha |E| \rfloor$ edges from $E$, where $0 < \alpha < 1$. Denote the set of removed edges as $E_s$, and the set of residual edges as $E_r$. Then, we compute the largest connected component $G_c = (V_c, E_c)$ of the graph induced on $E_r$. Denote $E_p$ as the set of edges in $E_s$ with nodes both in $V_c$. Finally, we can generate the training set as the edges in $E_c$, and the testing set consisting of two equal-sized parts, i.e., (i) the positive edges which are in $E_p$, and (ii) the negative edges which are pair of nodes $u$ and $v$ such that $(u, v)$ is not an edge in $E$. Note that, in the experiments, we set $\alpha = 10\%$ and the number of positive edges in the testing set equals the one of negative edges.

Table 2: Performance in link prediction evaluated by precisions with cosine similarity and euclidean similarity.

| Algorithm | Cosine Similarity | | | | | | | Euclidean Similarity | | | | | | |
|---|---|---|---|---|---|---|---|---|---|---|---|---|---|---|
| | Blog | Flickr | YT | SP | UK | Twitter | FS | Blog | Flickr | YT | SP | UK | Twitter | FS |
| node2vec | 0.9628 | 0.8624 | 0.9731 | - | - | - | - | 0.9580 | 0.8369 | 0.9514 | - | - | - | - |
| NetSMF | 0.9669 | 0.8512 | - | - | - | - | - | 0.9593 | 0.8172 | - | - | - | - | - |
| ProNE | 0.9716 | 0.8697 | 0.9745 | - | - | - | - | 0.9612 | 0.8436 | 0.9591 | - | - | - | - |
| SepNE | 0.9633 | **0.8710** | 0.9728 | 0.7313 | - | - | - | 0.9595 | **0.8493** | 0.9568 | 0.7246 | - | - | - |
| MILE | 0.9458 | 0.8307 | 0.9437 | 0.6823 | 0.8567 | 0.6117 | - | 0.9128 | 0.8071 | 0.9232 | 0.6483 | 0.8188 | 0.6359 | - |
| ParNE | 0.9657 | 0.8369 | 0.9518 | 0.7139 | 0.8473 | 0.6294 | - | 0.9628 | 0.8353 | 0.9408 | 0.6946 | 0.8519 | 0.6530 | - |
| PBG | 0.9614 | 0.8598 | 0.9746 | 0.7519 | 0.8708 | 0.6517 | 0.6381 | 0.9572 | 0.8379 | 0.9554 | 0.6819 | 0.8435 | 0.6417 | 0.6189 |
| DistNE$^{nb}$ ($k=50$) | 0.9527 | 0.8408 | 0.9653 | 0.7244 | 0.8521 | 0.6359 | 0.6283 | 0.9394 | 0.8217 | 0.9304 | 0.7036 | 0.8380 | 0.6311 | 0.6055 |
| DistNE ($k=100$) | 0.9691 | 0.8587 | 0.9802 | 0.7557 | 0.8723 | 0.6664 | 0.6418 | 0.9625 | 0.8482 | 0.9527 | 0.7299 | 0.8571 | 0.6589 | 0.6204 |
| DistNE ($k=50$) | **0.9754** | 0.8618 | **0.9813** | **0.7588** | **0.8792** | **0.6704** | **0.6476** | **0.9680** | 0.8456 | **0.9649** | **0.7369** | **0.8627** | **0.6690** | **0.6245** |

Based on that, we compute the network embedding on the training set for each dataset, and then calculate the similarity of all pairs of nodes in the testing set. Finally, we compute the precision as the fraction of positive edges in the most similar $|E_p|$ pairs of nodes in the testing set.

Table 2 presents the precisions of all algorithms on all datasets. Note that, in order to study the effect of $k$ in the performance of DistNE, we vary $k$ from 50 to 100. As we can see, when $k = 50$, DistNE outperforms the competitors in almost all the cases. In particular, on the *Twitter* dataset, DistNE improves PBG by 4.25% in euclidean similarity and 2.82% in cosine similarity, due to that DistNE utilizes induced subgraphs and border subgraphs to preserve the internal and external structural information that favors the prediction of close relations, resulting in the superior performance. Besides, when $k = 100$, the performance of DistNE is still better than most of the competitors, but it degrades slightly, since more partitions leads to more edge cuts that lose the connection between nodes. Moreover, when $k = 50$, we compare DistNE with the version without processing border subgraphs, denoted by DistNE$^{nb}$. As shown in Table 2, DistNE$^{nb}$ cannot provide satisfactory performance, since it ignores the external information on the border subgraphs.

### 5.4 Experiments on Node Classification

Then we evaluate the effectiveness of DistNE by performing the node classification tasks on 4 datasets, i.e., *Blog*, *Flickr*, *Youtube*, and *Spammers*, whose nodes have multiple labels. In particular, we run all algorithms on each dataset, and obtain the network embedding of each dataset. Besides, we randomly split the set of nodes with labels into two even-sized disjoint subsets, denoted by training and testing sets respectively. Afterwards, treating the network embedding as features of nodes, we build on the training set a multi-class logistic regression classifier that utilizes one-vs-rest technique and L2 regularization. Finally, following the previous work [12], we measure the performance of all algorithms in task of node classification on the testing set by micro-F1 and macro-F1 scores.

Table 3 provides the micro-F1 scores and macro-F1 scores of all competitors on the 4 datasets. As aforementioned, we vary $k$ from 50 to 100 for DistNE. Observe that, when $k = 50$, DistNE consistently outperforms the competitors on all datasets. In particular, on the *Flickr* dataset, DistNE is better than the second-best approach, i.e., SepNE, by 4.27% improvement in Macro-F1 score and 3.87% in Micro-F1 score. This is because DistNE separates the internal and external structural information of nodes, which empowers more discrimination. Besides, when $k = 50$, DistNE significantly outperforms DistNE$^{nb}$, which illustrates that the external information on the border subgraphs highly augments the quality of network embedding. Furthermore, when $k = 100$, while the performance of DistNE decreases slightly, it is still better than most of the competitors, which again demonstrates the superiority of DistNE.

## 6 DEPLOYMENT

We have deployed the proposed distributed algorithms in several online games of Tencent with various applications, as illustrated in the sequel.

### 6.1 Deployment Setup

In this paper, we present two different games, denoted by X and Y respectively. Game X is a multiplayer online battle royale game, and Game Y is a multiplayer online battle arena game. For each game, we construct its social graph by taking each player in the game as a node and the friendship between two players as an edge. Both graphs have several billions of edges, as shown in Table 4.

After that, we run the distributed network embedding algorithms on the social graphs of games X and Y respectively using the in-house cluster with 51 machines, as explained in Section 5. Then, we obtain the network embedding of players in each game, which are then used in the downstream applications, i.e., friend recommendation and item recommendation. In each application, we train the model for recommendation by taking the network embedding of players as their features. Besides, we compare the network embeddings produced by DistNE with the alternative approach, i.e., PBG, which is evaluated by the online A/B testing [31] that randomly assigns a fraction of live traffic to the treatment groups, i.e., the players receiving the recommendations from the different approaches. Besides, we update the network embedding of players every 7 days by re-running the algorithms on the latest social graphs, and report the average readings over 4 consecutive observation periods, each of which lasts 7 days.

Table 5 shows the running time of DistNE and PBG on both games respectively. DistNE is faster than PBG by 91.11% (resp. 60.56%) on Game X (resp. Game Y), due to the superior parallelism with subgraphs, as explained in Section 5.1.

### 6.2 Friend Recommendation

In the online games, a player might want to connect with the other players for the purpose of social to interact with interesting people, or gaming requirements that encourage players to play the games together. However, it is difficult for a player to search among billions or millions of players. Instead, we provide an in-game module to recommend at most 50 players that one would be interested in. When the player $v$ accesses the module in the games, $v$ sees a list of recommended players, on which $v$ can click one of them $u$ if interested. After that, the clicked player $u$ receives an invitation of making friends from $v$. Player $u$ accepts the invitation if interested, otherwise reject. As such, we evaluate the approach for friend recommendation by two metrics: (i) click rate, which is the fraction of players clicking the recommendations over the ones seeing the recommendations; and (ii) approval rate, equal to the fraction of players accepting the invitation over the ones receiving the invitations.

Table 3: Performance in node classification evaluated by F1 scores.

| Algorithm | Micro-F1 score | | | | Macro-F1 score | | | |
|---|---|---|---|---|---|---|---|---|
| | Blog | Flickr | YT | SP | Blog | Flickr | YT | SP |
| node2vec | 0.2922 | 0.1286 | 0.1952 | - | 0.2213 | 0.0834 | 0.1264 | - |
| NetSMF | 0.2883 | 0.1267 | - | - | 0.2177 | 0.0796 | - | - |
| ProNE | 0.2986 | 0.1353 | 0.2041 | - | 0.2269 | 0.0902 | 0.1291 | - |
| SepNE | 0.2951 | 0.1368 | 0.2149 | 0.4286 | 0.2229 | 0.0913 | 0.1358 | 0.3671 |
| MILE | 0.2831 | 0.1219 | 0.1859 | 0.4059 | 0.2134 | 0.0765 | 0.1173 | 0.3493 |
| ParNE | 0.2918 | 0.1240 | 0.2114 | 0.4133 | 0.2198 | 0.0782 | 0.1309 | 0.3556 |
| PBG | 0.2937 | 0.1311 | 0.1985 | 0.4174 | 0.2231 | 0.0860 | 0.1278 | 0.3582 |
| DistNE$^{nb}$ $_{(50)}$ | 0.2623 | 0.1234 | 0.1933 | 0.4064 | 0.2076 | 0.0742 | 0.1216 | 0.3489 |
| DistNE $_{(100)}$ | 0.2992 | 0.1403 | 0.2156 | 0.4311 | 0.2285 | 0.0920 | 0.1366 | 0.3731 |
| DistNE $_{(50)}$ | **0.3026** | **0.1421** | **0.2170** | **0.4355** | **0.2317** | **0.0952** | **0.1394** | **0.3789** |

Table 4: The graphs in the deployed games of Tencent.

| Game | Type | #Nodes | #Edges |
|---|---|---|---|
| X | Shooting | 0.27 billion | 8.54 billion |
| Y | MOBA | 0.76 billion | 20.58 billion |

Table 5: Running time on the graphs of games.

| Algorithm | Game X | Game Y |
|---|---|---|
| PBG | 25.8 hours | 51.7 hours |
| DistNE | **13.5 hours** | **32.2 hours** |

Based on the network embedding computed by DistNE and PBG respectively, for each player $v$, we utilize the locality sensitive hashing [18] to compute the players $u$ who are not friends of $v$ and have the embedding $f(u)$ among the top-50 closest distance to player $v$'s embedding $f(v)$. In the end, we recommend the top-50 closest players to player $v$.

Table 6 illustrates the performance of PBG and DistNE for friend recommendation in games X and Y respectively. As shown, DistNE is consistently better than PBG, since DistNE employs the recursive graph partition that captures the local structural information of the graph, which favors the close and well connected relations. Specifically, in game X (resp. Y), DistNE outperforms PBG by 6.46% (resp. 6.62%) on click rate and by 12.80% (resp. 3.70%) on approval rate.

### 6.3 Item Recommendation

Another application of DistNE in the games is the item recommendation, where we are to recommend each player a list of items to purchase in the in-game shop. To achieve that, we utilize the machine learning methods to learn the preference of players to the items. In particular, we first extract the features of players by DistNE on the graph, and also from the gaming data, such as demographics and game profiles. Besides, we generate the features of items from the purchasing data. Then, we exploit the purchasing and viewing data to generate the labels for the pairs of player $v$ and item $i$. That is, we label $(u, i)$ as positive if $u$ purchased $i$, and negative if $u$ saw $i$ but did not purchase $i$. Based on the positive and negative labels, as well as the features of players and items, we train a binary classifier using XGBoost [4], which is thereafter used to predict the probability that a player purchases a given item.

Table 7 compares the purchase rate for the models using PBG and DistNE in the games X and Y respectively. As we can see, DistNE outperforms PBG by 5.21% (resp. 3.78%) in game X (resp. Y), which demonstrates the effectiveness of induced subgraphs and border subgraphs that well preserve the internal and external structural information respectively.

## 7 OTHER RELATED WORK

Besides the work introduced in Section 1 and Section 2.2, there are some other work aiming at accelerating the generation of network embedding on a single machine, which can be roughly classified into two categories, as follows. One category of these work, such as MILE [21] and GraphZoom [7], coarsens the graph recursively in several iterations, each of which halves the size of graph, and computes the network embedding for the smallest coarsened graph, which are used to generate the embeddings of the input graph. However, due to the recursive computation, this line of work is difficult to be translated in parallel. Another category of these work exploits the sparsification or the separation of graph by matrix manipulations to reduce the computational cost, such as Progle [38], NetSMF [29], ProNE [37], and SepNE [20]. However, the matrix manipulations are costly for large graphs, especially when the size of graph has already exploded the memory space of a single machine. Recently, Lin et al. [23] devise an efficient and effective method for the initialization of network embedding algorithms, which utilizes the graph partition technique. However, this method targets at the quality of initialization, rendering it insufficient for the ultimate goal of network embedding. Moreover, there are some work [36] speedup the processing of graph neural network by mutli-threading technique, which cannot be directly translated in distributed computing for the problem of network embedding.

Furthermore, to consider the locality of graph structure, some work [8, 10, 25] incorporate the community information in the network embedding, or generate multiple embeddings with respect to different local structures. However, these approaches do not take into consideration the metrics of distributed computing, e.g., load balancing and communication minimization.

On the other hand, there exist some work for computing random walks on large graphs [22, 34], which can be used to generate the training samples in the network embedding algorithms. However, they cannot solve the massive cost of model training, which would require the exchange of data between nodes on the graph. The other line of research [14] focuses on the inductive network embedding by sampling and aggregating neighbors, which is orthogonal to the problem of this paper, i.e., the transductive one.

## 8 CONCLUSIONS

In this paper, we present DistNE as an efficient and effective distributed algorithm for network embedding on large graphs. We devise the recursive graph partitioning technique that divides the graph into sufficiently small subgraphs by considering the size of available memory and the number of border nodes. As such, the subgraphs can well preserve the internal and external structural information of nodes. Then, the network embedding of all subgraphs are computed independently in parallel, and aggregated with a linear cost to generate the final embeddings. In various experiments, we demonstrate that DistNE is faster than the state-of-the-art approaches by several times and outperforms the competitors in the tasks of link prediction and node classification. Finally, we deploy DistNE in the applications of two games of Tencent respectively, and show that DistNE improves the baselines by a large fraction in the evaluation metrics.

Table 6: Performance in friend recommendation.

| Algorithm | Game X | | Game Y | |
|---|---|---|---|---|
| | Click Rate | Approval Rate | Click Rate | Approval Rate |
| PBG | 0.2042 | 0.1148 | 0.3216 | 0.7033 |
| DistNE | **0.2174** | **0.1295** | **0.3429** | **0.7293** |

Table 7: Purchase rate in item recommendation.

| Algorithm | Game X | Game Y |
|---|---|---|
| PBG | 0.0653 | 0.2381 |
| DistNE | **0.0687** | **0.2471** |